\documentstyle[12pt]{article}
\title{The trivial solution of the gravitational energy-momentum tensor problem}
\author{Hrvoje Nikoli\'c \\
Theoretical Physics Division, Rudjer Bo\v{s}kovi\'{c} Institute, \\
P.O.B. 180, HR-10002 Zagreb, Croatia \\
{\normalsize e-mail: hnikolic@irb.hr} \\
\makebox[1in]{} \\
}
\date{\today}
\begin{document}
\maketitle
\begin{abstract}
In the literature one often finds the claim that there is no such thing
as an energy-momentum tensor for the gravitational field, and consequently,
that the total energy-momentum conservation can only be defined in terms 
of a gravitational energy-momentum pseudo-tensor.
Nevertheless,  by relaxing the assumption that gravitational energy-momentum tensor
should only depend on first derivatives of the metric,
the Einstein equation leads to a trivial result that gravitational energy-momentum tensor is essentially
the Einstein tensor.
We discuss various peculiarities of such a definition of energy-momentum are argue 
that all these peculiarities have a sensible physical interpretation.
\end{abstract}
\vspace*{0.5cm}
%04.20.-q 	Classical general relativity 
%04.20.Cv 	Fundamental problems and general formalism

\section{Introduction}

In general relativity, the matter energy-momentum tensor $T^{\mu\nu}$ satisfies
the covariant conservation law 
\begin{equation}\label{e1}
 \nabla_{\mu}T^{\mu\nu} =0.
\end{equation}
Unlike the local covariant conservation $\nabla_{\mu}j^{\mu}=0$
of a vector $j^{\mu}$, the local covariant conservation (\ref{e1}) of a tensor,
in general, does not lead to a global conservation of matter energy.
If $n^{\mu}$ is the unit vector normal to a spacelike hypersurface $\Sigma$,
the global matter energy $\int_{\Sigma} d^3x \sqrt{|g^{(3)}|} n^{\mu} n^{\nu}T_{\mu\nu}$,
in general, depends on $\Sigma$. (An exception is a spacetime with a symmetry characterized
by a timelike Killing vector $\xi^{\mu}$, because then one can
introduce the local energy-momentum {\it vector} $p^{\mu}=T^{\mu\nu}\xi_{\nu}$. In this case 
(\ref{e1}) implies the local vector conservation $\nabla_{\mu}p^{\mu}=0$, 
which follows from the facts that (i) $T^{\mu\nu}$ is a symmetric tensor and (ii)
the Killing vector $\xi^{\mu}$ obeys $\nabla_{\mu}\xi_{\nu}+\nabla_{\nu}\xi_{\mu}=0$.)

The fact that (\ref{e1}) does not imply global conservation of matter energy
has a simple physical interpretation: the energy-momentum of matter can be exchanged
with the energy-momentum of the gravitational field. But this suggests that there should
be an energy-momentum tensor $t^{\mu\nu}$ of the gravitational field itself, such that the
total energy-momentum tensor
\begin{equation}\label{e2}
 T^{\mu\nu}_{\rm tot}=T^{\mu\nu}+ t^{\mu\nu}
\end{equation}
is conserved in the ordinary sense
\begin{equation}\label{e3}
\partial_{\mu}T^{\mu\nu}_{\rm tot}=0. 
\end{equation}
If so, then one can introduce the global 4-momentum
\begin{equation}\label{e4}
 P^{\mu}_{\rm tot}=\int d^3x\, T^{\mu 0}_{\rm tot}, 
\end{equation}
which, due to (\ref{e3}), obeys the global conservation
\begin{equation}\label{e5}
 \frac{dP^{\mu}_{\rm tot}}{dx^0}=0.
\end{equation}

Nevertheless, in general-relativity textbooks one often finds the claim
that such a gravitational energy-momentum tensor $t^{\mu\nu}$ does not exist
\cite{eddington,bergmann,weyl,pauli,fock,LL,weinberg,MTW,dirac,ABS,dFC,stephani,
HEL,ryder,padmanabhan,zee,PW}.
According to the mentioned textbooks, the best one can do is to construct 
a pseudo-tensor quantity $t^{\mu\nu}$ which does not transform as a tensor 
under general coordinate transformations.
The pseudo-tensor $t^{\mu\nu}$ can be chosen in many inequivalent ways \cite{padmanabhan},
while the most popular choice is the one by Landau and Lifshitz \cite{LL}.

Contrary to this widely accepted claim that the gravitational energy-momentum tensor $t^{\mu\nu}$ 
does not exist, in this paper we point out that it does. 
Moreover, it turns out to be trivial to construct it, if one is willing to relax one common 
assumption -- that $t^{\mu\nu}$ should be constructed from the metric $g_{\mu\nu}$ and its
first derivatives $\partial_{\alpha}g_{\mu\nu}$. By allowing $t^{\mu\nu}$ to depend also
on the second derivatives $\partial_{\alpha}\partial_{\beta}g_{\mu\nu}$, we find the trivial
solution to the problem of constructing the gravitational energy-momentum tensor $t^{\mu\nu}$;
the appropriate tensor $t^{\mu\nu}$ turns out to be proportional to the Einstein tensor
$G^{\mu\nu}$.

Indeed, such a definition of energy-momentum has also been proposed a long time ago 
by Lorentz \cite{lorentz} and Levi-Civita \cite{levi-civita}. However, textbooks
rarely mention the possibility of such a definition of energy-momentum, 
and when they do, they dismiss it as
inadequate \cite{pauli,zee,blau}. In this paper we reexamine various peculiarities 
of such a definition of energy-momentum and argue that these peculiarities
are not a valid reason to dismiss it.

\section{The gravitational energy-momentum tensor}

Let us start from the Einstein equation
\begin{equation}\label{e6}
 G^{\mu\nu}=8\pi G_{\rm N}T^{\mu\nu},
\end{equation}
where $G_{\rm N}$ is the Newton constant and $G^{\mu\nu}$ is the Einstein tensor
\begin{equation}\label{e7}
 G^{\mu\nu} \equiv R^{\mu\nu}-\frac{1}{2}g^{\mu\nu}R. 
\end{equation}
Eq.~(\ref{e6}) can also be written as
\begin{equation}\label{e8}
 T^{\mu\nu}-\frac{1}{8\pi G_{\rm N}}G^{\mu\nu}=0 .
\end{equation}
Applying the derivative $\partial_{\mu}$ on both sides of (\ref{e8}), one gets
\begin{equation}\label{e9}
 \partial_{\mu} \left( T^{\mu\nu}-\frac{1}{8\pi G_{\rm N}}G^{\mu\nu} \right)=0 .
\end{equation}
But this is precisely the ordinary conservation equation (\ref{e3}), 
provided that in (\ref{e2}) one makes the identification 
\begin{equation}\label{e10}
 t^{\mu\nu}\equiv -\frac{1}{8\pi G_{\rm N}}G^{\mu\nu} .
\end{equation}
Hence, the tensor (\ref{e10}) can naturally be interpreted as the energy-momentum tensor
of the gravitational field. It depends on the metric $g_{\mu\nu}$ and its
first and second derivatives $\partial_{\alpha}g_{\mu\nu}$ and 
$\partial_{\alpha}\partial_{\beta}g_{\mu\nu}$, respectively. 

The matter energy-momentum tensor $T^{\mu\nu}$ usually depends on matter fields and their
first derivatives, but not on second derivatives of the matter fields. 
Nevertheless, there is no any physical reason why it should be the case for all 
energy-momentum tensors. Therefore we do not see any physical problem with the fact
that the gravitational energy-momentum tensor (\ref{e10}) depends on the second derivatives 
of the gravitational field $g_{\mu\nu}$.
 
That (\ref{e10}) is the natural energy-momentum tensor for the gravitational field 
can also be seen from the total action 
\begin{equation}\label{e11}
 S_{\rm tot}=S_{\rm grav} + S_{\rm matter} ,
\end{equation}
where $S_{\rm matter}$ is the matter action and $S_{\rm grav}$ is the pure gravity action
\begin{equation}\label{e12}
 S_{\rm grav}=\frac{1}{16\pi G_{\rm N}} \int d^4x \sqrt{|g|} R .
\end{equation}
The matter energy-momentum tensor is defined as (see e.g. \cite{carroll})
\begin{equation}\label{e13}
 T_{\mu\nu}=\frac{-2}{\sqrt{|g|}} \frac{\delta S_{\rm matter}}{\delta g^{\mu\nu}} .
\end{equation}
Likewise, by defining the gravitational energy-momentum tensor as
\begin{equation}\label{e14}
 t_{\mu\nu}=\frac{-2}{\sqrt{|g|}} \frac{\delta S_{\rm grav}}{\delta g^{\mu\nu}} ,
\end{equation}
one recovers (\ref{e10}). In the same spirit, one can define the total
energy momentum tensor as
\begin{equation}\label{e15}
 T_{\mu\nu}^{\rm tot}=\frac{-2}{\sqrt{|g|}} \frac{\delta S_{\rm tot}}{\delta g^{\mu\nu}} ,
\end{equation}
which leads to
\begin{equation}\label{e16}
 T^{\mu\nu}_{\rm tot}=T^{\mu\nu}-\frac{1}{8\pi G_{\rm N}}G^{\mu\nu} .
\end{equation}
In this way the Einstein equation (\ref{e8}) can be interpreted as a constraint
that the total energy-momentum tensor must vanish. 

The vanishing of the total energy-momentum tensor is the main source of the critique
of (\ref{e16})
in the literature \cite{pauli,zee,blau} . Essentially, it is claimed that a concept 
of a vanishing energy-momentum is useless. While we agree that a vanishing 
energy-momentum is less useful than energy-momentum which can take different values
in different physical situations, we do not accept that it is totally useless. 
In particular, vanishing of the total energy-momentum tensor
can also be viewed
as a covariant version of the Hamiltonian constraint ${\cal H}_{\rm tot}=0$ appearing
in the canonical formulation of gravity \cite{MTW,padmanabhan,bojowald}.
In the quantum theory, the vanishing of the total Hamiltonian has a very deep physical 
consequence, leading to the famous problem of time in quantum gravity \cite{kuchar,isham}.
To note at least one possible use of it, let us only mention that it might be a key
to the solution of the black-hole information paradox \cite{nik_time_loc_obs}.   

Zee \cite{zee} makes a further critique of a vanishing total energy-momentum
by comparing it with the Newton equation written as $F-ma=0$, which one might attempt to 
interpret as the claim that ``the total force vanishes''. 
While there is some point in such a comparison, in our opinion it misses the deeper 
geometrical message of the Einstein equation, which expresses the fact that general relativity
is {\em diffeomorphism invariant}. In particular, it means that $\mu 0$ and $0\nu$ components
of the Einstein equation (\ref{e6}) are not really analogous to the Newton equation, 
but are non-dynamical constraint equations not containing second time derivatives.
In this sense, a better analogue is a classical particle with an action invariant
under reparameterizations of the time coordinate, leading to the vanishing total
Hamiltonian (see e.g. \cite{padm2,nik_covar_can}). 

Another related unappealing feature of such a definition of the gravitational energy-momentum
$t^{\mu\nu}$ is the fact that
it vanishes at all points at which $T^{\mu\nu}$ vanishes. In particular, it means 
that the gravitational wave propagating through a spacetime without matter carries zero 
energy-momentum. But this result should not be surprising, given that the very definition
of the gravitational wave is non-covariant in essence. Namely, 
the definition of gravitational waves rests on a non-covariant
split of the metric $g_{\mu\nu}=\gamma_{\mu\nu}+h_{\mu\nu}$, where $\gamma_{\mu\nu}$ is 
an arbitrary background metric (usually chosen to be the Minkowski metric $\eta_{\mu\nu}$)
and $h_{\mu\nu}$ is a disturbance, the propagation of which is identified with the gravitational wave. 
Neither $\gamma_{\mu\nu}$ nor $h_{\mu\nu}$ transforms as a tensor under general coordinate transformations.
Thus the fact that the covariant energy-momentum tensor of a gravitational wave can vanish reflects
the fact that the gravitational wave itself is not a covariant object.

Note also that Einstein equation (\ref{e8}) and positivity of the matter energy-density $T^{00}$ 
imply that gravitational energy-density $t^{00}$ is negative at points at which matter is present.
This negativity of gravitational energy reflects the attractive nature of gravity when it acts on 
matter.

A possible reason for worry is also the fact that the left-hand side of (\ref{e9}) is not a tensor,
owing to the fact that the ordinary derivative $\partial_{\mu}$ is not a covariant object.
In most cases that would be a problem, but here it is not a problem because (\ref{e9}) is valid 
in all coordinate frames. This is a consequence of the fact that the bracket 
in (\ref{e9}) vanishes itself due to (\ref{e8}), so that the vanishing of the derivative
of the bracket is rather trivial.  

%The final objection against the discussed definition of the gravitational energy-momentum
%tensor might be its triviality. But even if the whole idea looks trivial once it is exposed,
%the triviality is not necessarily a drawback.  
%Such a definition of gravitational energy-momentum does not seem to be widely known in the community, 
%so we believe that our explicit exposition of the idea can be illuminating. 

To conclude, we believe that there are good arguments for accepting the gravitational
energy-momentum tensor (\ref{e10}) as physically viable, despite of some peculiarities
associated with it.

\section*{Acknowledgments}

The author is grateful to M. Blau and J. Pereira for useful discussions 
and for drawing attention to some relevant references.  
This work was supported by the Ministry of Science of the
Republic of Croatia.

\end{document}